\documentstyle[11pt,newpasp,twoside,epsf]{article} 
\def\edcomment#1{\iffalse\marginpar{\raggedright\sl#1\/}\else\relax\fi} 
\reversemarginpar

\begin{document} 

\title{Pair Creation at Shocks: Application to the High Energy
Emission of Compact objects}
\author{P.O. Petrucci}
\affil{Osservatorio Astronomico di Brera, Milano, Italy}
\author{G. Henri$^2$, G. Pelletier$^2$} 
\affil{Laboratoire d'Astrophysique, Grenoble, France}

\begin{abstract} 
We investigate the effect of pair creation on a shock
structure. Actually, particles accelerated by a shock can be
sufficiently energetic to boost, via Inverse Compton (IC) process for
example, surrounding soft photons above the rest mass electron energy
and thus to trigger the pair creation process. The increase of the
associated pair pressure is thus able to disrupt the plasma flow and
possibly, for too high pressure, to smooth it completely. Reversely,
significant changes of the flow velocity profile may modify the
distribution function of the accelerated particles, modifying
consequently the pair creation rate. Stationary states are then
obtained by solving self-consistently for the particle distribution
function and the flow velocity profile. We discuss our results and the
application of these processes to the high energy emission and
variability of compact objects.
\end{abstract} 

\section{Introduction}
The high energy emission observed in compact objects like AGNs or
X-ray binaries requires the existence of high energy particles. Shocks
being particularly attractive particle acceleration sites, they are
generally believed to occur in the central region of these objects. In
these cases however, the presence of important radiation fields
necessitate to take into account particles-photons interactions in the
shock region.\\ If some works have already studied particles
acceleration at shocks including radiative cooling processes (Webb et
al., 1984; Drury et al., 1999) we have investigated the effect of pair
creation on the shock structure (Petrucci et al., hereafter
P00). Particles accelerated by the shock can effectively be
sufficiently energetic to boost, via Inverse Compton (IC) process for
example, surrounding soft photons above the rest mass electron energy
and thus enable to generate pairs. Consequently, the increase of the
pair pressure may be able to modify the plasma flow and eventually,
for too high pressure, to smooth it completely.\\ We describe here the
geometry of the (toy) model we use (cf. Fig 1). We present the basic
equations and the main results of this model. We then briefly discuss
its application to the high energy emission and variability of compact
objects.

\section{The Toy Model}
The schematic view of our toy model is plotted in Fig. 1.  We suppose
the existence of a thermal supersonic plasma undergoing an adiabatic
{\bf non-relativistic shock}. It is supposed to {\bf be embedded in an
isotropic external soft photon field}. We assume the presence of a
magnetic field $\vec{B}_o$, parallel to the shock normal, frozen in
the plasma and slightly perturbed by Alfven waves. We suppose the
magnetic energy to be in equipartition with the particles kinetic
energy so that, in each part of the shock, the Alfven wave speed is
equal to the flow speed. Particles are then scattered by Alfven waves
through pitch angle scatterings, gaining energy through the {\bf first
order Fermi process} when crossing back and forth the shock
discontinuity. We also suppose the magnetic perturbations to have
sufficiently small amplitudes so that we can treat the problem in
quasilinear theory using the {\bf Fokker-Planck formalism}. Finally,
we suppose the {\bf plasma pressure to be dominated by the pressure of
the relativistic particles}. In this case, the flow velocity profile
makes already a smooth transition between the up and downstream region
but acceleration still occurs.

\subsection{The Geometry}
\subsubsection{The shock location:}
We assume a 1D geometry. It means that the different parameters
characterizing the flow are supposed to be homogeneous in each section
perpendicular to the $x$ axis. This is a relatively good approximation
in the central parts of the shock where the border effects are
negligible. We suppose the shock to be located in $x=0$
(cf. Fig. 1). During numerical integrations, this will be ensuring by
imposing that the flow velocity $u(x)$ possesses an inflection point
in $x=0$ that is:
\begin{equation}
\left.\frac{\partial^2 u}{\partial
x^2}\right|_{x=0}=0
\end{equation}
\begin{figure}[h!]
\plotone{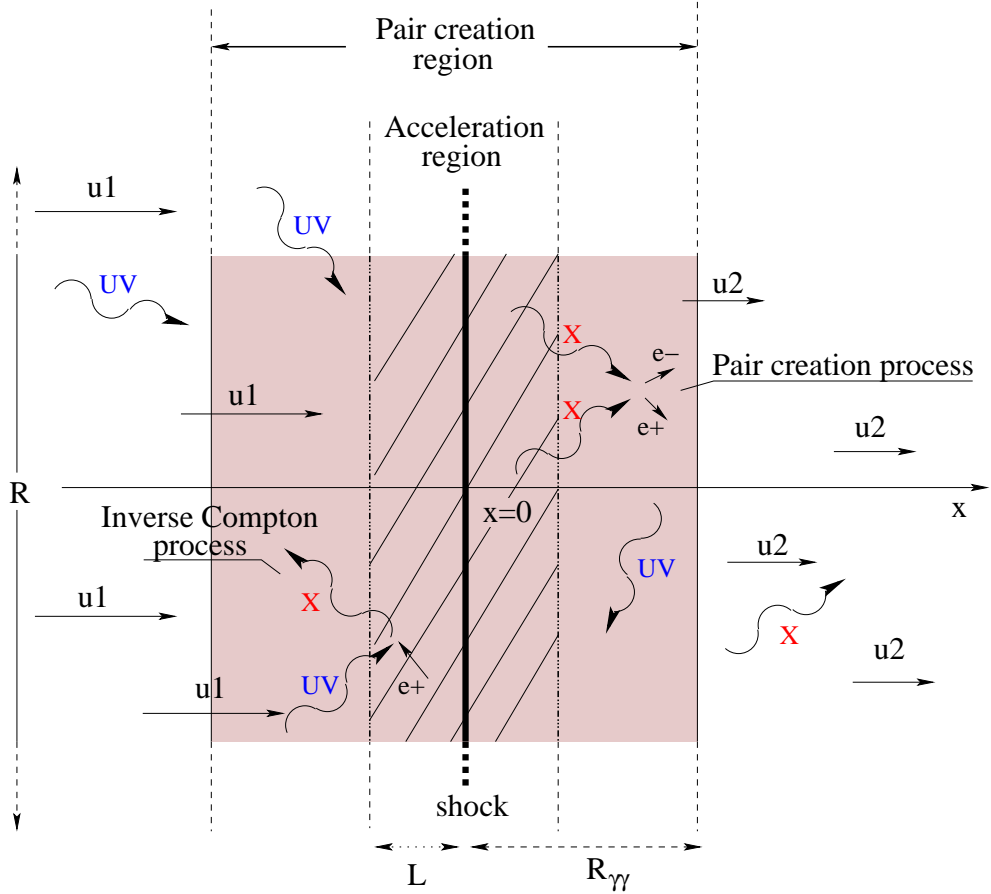}
\caption{The geometry of our (toy) model. The shock discontinuity is
represented by the bold line. The hatched region is the acceleration
region and the pair creation one is filled in pink. Particles are
represented by straight arrows and photons by warped ones (in blue for
surrounding soft, i.e. UV, photons, and in red, high energy,
i.e. X-ray, photons). Scales are not respected.}
\end{figure}

\subsubsection{The acceleration region:}
A particle of Lorentz factor $\gamma$ will interact with the shock if
it is located within about one diffusion length
$L_{D}(\gamma)=D_{xx}/u$ (where $D_{xx}$ is the spatial diffusion
coefficient) from the shock. On the other hand it will be cooled (by
Inverse Compton process) on a cooling length scale
$L_{cool}(\gamma)$. $L_{D}$ and $L_{cool}$ are increasing and
decreasing functions of $\gamma$ respectively. There thus exists a
Lorentz factor $\gamma_c$ for which:
\begin{displaymath} 
L_{D}(\gamma_c)=L_{cool}(\gamma_c)=L
\end{displaymath}
We will define the acceleration region as the physical space $-L\le
x\le L$ (hatched region in cf. Fig. 1). Consequently, in this region
the {\bf coolings are negligible in comparison to heatings} for
particles with $\gamma< \gamma_c$ since for such particles
$L_{D}(\gamma)< L_{cool}(\gamma)$.\\

In the following we will suppose the spatial diffusion coefficient
$D_{xx}$ to be independent of the energy of the particles and it will
be simply written $D$.

\subsubsection{The pair creation region:}
Particles accelerated in the shock will produce high energy photons by
scattering, via Inverse Compton process (IC), the external soft
photons. These high energy photons will interact with themselves to
produce pairs on a length scale $R_{\gamma\gamma}$. Typically we have:
\begin{displaymath}
R_{\gamma\gamma}\simeq\frac{R}{1+\tau_{\gamma\gamma}}
\end{displaymath}
where $R$ is the typical size of the shock (cf. Fig. 1). This
expression take into account the photon-photon depletion for
$\tau_{\gamma\gamma}\gg$ 1 and the geometrical dilution of the X-ray
photon density, far from the shock, which upperlimits the pair
creation region size to roughly the shock size $R$. We will define the
pair creation region by $-R_{\gamma\gamma}\le x\le
+R_{\gamma\gamma}$.\\

We assume the pair pressure creation rate to be constant in the pair
creation region (that is for $-R_{\gamma\gamma}\le x\le
R_{\gamma\gamma}$) and null outside. This assumption may be supported
by the fact that magnetic turbulence, supposed to be present to
scatter particles, is a useful process to isotropize and homogenize
the plasma in the vicinity of the shock in very small time scale

\subsubsection{The ``effective'' compression ratio:}
The definition of the compression ratio like the ratio between the far
upstream and downstream flow velocity is rather unsatisfactory in our
case since it will not give a real estimate of the velocity change
experienced by a relativistic particle in the vicinity of the shock,
where the particle distribution is principally built. We will thus
define an ``effective'' compression ratio, noted simply $r$, as the
ratio between the upstream velocity in $-10L$ and the downstream
velocity in $+10L$. We have checked that in the case of a strong shock
without pairs, this definition still gives a compression ratio very
near the expected value (in a plasma dominated by relativistic
particles) of 7.

\subsection{Important energetic al thresholds}
\subsubsection{The acceleration threshold:}
Only particles having a Larmor radius $r_L$ comparable to the
wavelength of the Alfven spectrum will undergo scatterings (Jokipii,
1976; Lacombe, 1977) and thus will go back and through many times
across the shock.  In $e^--p^+$ plasma as those we deal with, the
non-relativistic protons limit the Alfven waves spectrum to wavelength
greater than $2\pi V_A/\omega_{cp}$ ( $\omega_{cp}$ is the cyclotron
pulsation of the protons in a magnetic field B, and $V_A$ is the
Alfven velocity). Consequently there exists a lower Lorentz factor for
a relativistic lepton to be accelerated in a shock:
\begin{displaymath} 
\gamma>\gamma_{min}=\frac{m_p}{m_e}\frac{V_A}{c}
\end{displaymath}
In the more general case $\gamma_{min}$ is at least of the order of a
few. We will admit that pre-accelerator processes exist (like magnetic
reconnection, whistler) to bring particles above $\gamma_{min}$.
Consequently, since particles annihilate preferentially for
$\gamma\simeq 1$ (Coppi \& Blandford, 1990), {\bf we will neglect the
annihilation process in the shock region}.

\subsubsection{The pair creation threshold:}
We will suppose for simplicity that the external soft photon field is
mono-energetic with a mean photon energy $\epsilon_s$ (in $m_ec^2$
unit). Thus, on average, a soft photon scattered by a lepton of
Lorentz factor $\gamma$ will be boosted, via IC, to an energy
$\displaystyle \epsilon\simeq \frac{4}{3}\gamma^2\epsilon_s$ (Rybicki
\& Lightman, 1979). The high energy photon produced will be able to
generate a pair electron-positron if at least:
\begin{displaymath} 
\epsilon > 2m_ec^2 \mbox{ i.e. }
\gamma\ge\left(\frac{3m_ec^2}{2\epsilon_s} \right)^{1/2}=\gamma_{th}
\end{displaymath}
We assume $\gamma_{th}>\gamma_{min}$ so that particles have to be
accelerated in the shock to initiate the pair creation process.

\subsection{The particle distribution}
\label{partdist}
\subsubsection{The spectral index:}
It can be shown that the solution of the evolution equation of the
particle distribution function including the pair creation process
(cf. Eq. (5)), still has a energy power law dependence, as it is
effectively the case without pairs (P00). The spectral index $s$ (here
$s$ is the spectral index of the spatially integrated distribution
function of the particles $n(\gamma)\propto\gamma^{-s}$) keeps also
the same expression in function of the compression ratio i.e.:
\begin{equation}
 s(r)=(r+2)/(r-1).
\label{eqs}
\end{equation}
For plasma dominated by relativistic pressure, the compression ratio
is necessarily smaller than 7 so that $s(r)$ is larger than 1.5.

\subsubsection{The high energy cut-off:}
\label{highcut}
Since we take into accounts the cooling, a high energy cut-off in the
particle distribution must necessarily appear at a Lorentz factor
$\gamma_c$ where heating and cooling balance (Webb et al.,
1984). Since we suppose that particles cool via inverse Compton
process (assuming that the external soft photon density is homogeneous
in the shock region) and that they are accelerated by the first order
Fermi process, the maximum Lorentz factor $\gamma_c$ achievable by the
acceleration process may be written as follows (P00):
\begin{equation}
\gamma_c\propto\frac{1}{l_s}\frac{u_1}{c}\frac{r-1}{3(1+r^2)}
\label{GAM}
\end{equation}
where $l_s$ is the soft compactness ($\displaystyle
l_s=\frac{L_{soft}\sigma_T}{4\pi Rm_ec^3}$, $m_e$ being the electron
mass and $L_{soft}$ the external soft radiation luminosity) and $u_1$
the upstream flow velocity.\\

From these different remarks, we will assume that the particle
distribution has a cut-off power law shape, i.e.:
\begin{equation}
n(\gamma)\propto \gamma^{-s}\exp\left(-\frac{\gamma}{\gamma_c}\right)
\label{eqno}
\end{equation}

\section{The basic kinetic equations}
As previously said, we suppose the existence of a magnetic field
$\vec{B}_o$, perpendicular to the shock front and slightly perturbed
by Alfven waves. The particles are thus scattered by these waves
trough pitch angle scattering (Jokipii, 1976; Lacombe, 1977) and can
cross the shock front several times before escaping unless they are
rapidly cooled by radiative processes.  During these scatterings, the
particle gain energy trough the well known first order Fermi
process. We also suppose the magnetic perturbations to have
sufficiently small amplitudes so that we can treat the problem in
quasilinear theory using the Fokker-Planck formalism. Besides we
assume that, in each part of the shock front, the scattering is
sufficient for the particle distribution to be nearly isotropic. With
these different assumptions, and when first order Fermi process just
as radiative losses and pair creation/annihilation are taken into
account, the particles distribution function $f(p,x)$ must verify the
following equation :
\begin{eqnarray}
  \frac{\partial f}{\partial t}+u\frac{\partial f}{\partial x} &=&
  \frac{1}{3}\frac{\partial u}{\partial x} p \frac{\partial
  f}{\partial p}+\frac{1}{p^2}\frac{\partial bp^4f}{\partial
  p}+\frac{\partial }{\partial x}D\frac{\partial f}{\partial x}
  \nonumber\\ & & +{\bf C}_{\pm}(f)+{\bf D}_{\pm}(f).  \label{eqfb}
\end{eqnarray}
The three first terms of the right member correspond to the first
order process, the radiative losses ($b$ being $>$ 0, the detailed
expression of $b$ in the case of Inverse Compton cooling can be found
in P00) and the spatial diffusion respectively, ${\bf C}_{\pm}(f)$ is
the pair creation rate and ${\bf D}_{\pm}(f)$ the annihilation one.
This equation is essentially the equation of the cosmic-ray transport
originally given by Parker (1965), Skilling (1975 and references
therein) except for the addition of the radiative losses and the pair
processes.\\

Since we suppose that the shocked plasma pressure is dominated by the
pressure of the relativistic particles, we can deduce, from Eq. 5, the
hydrodynamic equation linking pairs (through their pressure) and the
flow velocity $u(x)$, that is (in stationary state):
\begin{equation}
u\frac{\partial P_{\mbox{{rel}}}}{\partial x}+\frac{4}{3}
P_{\mbox{{rel}}}\frac{\partial u}{\partial x}=\frac{\partial
}{\partial x}D_{xx}\frac{\partial }{\partial x}P_{\mbox{{rel}}}+
\dot{Q}_{\mbox{{rel}}}+\dot{P}_{\pm}
\end{equation}
where $\dot{Q}_{\mbox{{rel}}}$ is the pressure loss rate due to
radiative losses and $\dot{P}_{\pm}$ the pair pressure creation
rate.\\ 

A way to solve Eq. (6) is to integrate it between $-\infty$ and
$+L$. In this case, we can neglect the cooling during the
integration. Indeed, even if for $x\le -L$ particles do not interact
with the shock, we have assumed that some processes apply and balance
coolings so that particles are injected above the resonance threshold
$\gamma_{min}$. On the other hand, in the shock region (that is $-L\le
x\le L$), we have seen in section 2.1 that the coolings are
negligible, in comparison to heatings, for particles with a Lorentz
factor smaller than $\gamma_c$ (the case of the majority of the
particles). Besides, we can also neglect the annihilation since it
occurs mainly at low energy, i.e. for particles with Lorentz factor
$\gamma\simeq 1$ (cf. Coppi \& Blandford, 1990), whereas we assume
$\gamma\ge\gamma_{min}$.  In these conditions, and with the assumption
that the pair creation is homogeneous in the region
$-R_{\gamma\gamma}<x<+R_{\gamma\gamma}$ (meaning $\dot{P}_{\pm}$
roughly constant) and null outside (meaning $\dot{P}_{\pm}$=0), the
integration of Eq. (6) between $-\infty$ and $L$, combined with the
momentum conservation equation, gives (in reduced units):
\begin{equation}
\frac{\partial \widetilde{u}}{\partial
      \widetilde{x}}=\frac{7}{6}(1-\widetilde{u})\left(\frac{1}{7}-
      \widetilde{u}\right)
      +\Pi\left(1+\frac{\widetilde{x}}{\widetilde{R}_{\gamma\gamma}}\right)
\end{equation}
where the reduced variables are defined as followed:
\begin{eqnarray}
  \widetilde{u} &=& \frac{u}{u_1}\\
  \widetilde{x} &=& \frac{x}{L} \mbox{ and }
  \widetilde{R}_{\gamma\gamma}= \frac{R_{\gamma\gamma}}{L}\\
  \Pi &=& \frac{\dot{P_{\pm}}R_{\gamma\gamma}}{\rho
u_1^3}.
\end{eqnarray}
$\Pi$ is the ratio of the pair luminosity (the pair power density
integrated on the ``1D'' volume $R_{\gamma\gamma}$) to the kinetic
energy flux of the flow. This parameter will play an important role in
the evolution of the shock profile as we will see in the following. We
can anticipate that a large value of $\Pi$ will be certainly
unfavorable to the formation of the shock. It means that all the
kinetic energy of the upstream flow will be dissipated in pair
creation processes.

\section{Shock disappearance}
\begin{figure}[h!]
\plotfiddle{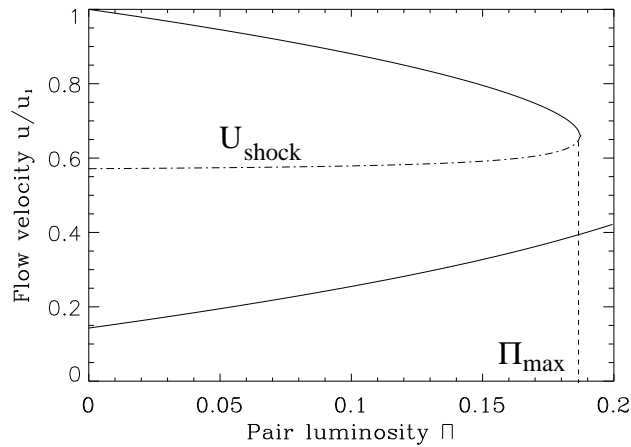}{6cm}{0}{50}{50}{-150}{-100}
\caption{Plot of the 3 solutions of Eq. (11) versus $\Pi$ for
$\widetilde{R}_{\gamma\gamma}=50$. The flow velocity
$\widetilde{u}_{\rm{schock}}$ at the shock location, i.e. in $x=0$ is
plotted in dashed line. There is a maximal value of $\Pi$,
$\Pi_{max}$, above which only one real solution exists. Consequently,
for $\Pi >\Pi_{max}$ the shock disappears.\label{a_urelation}}
\end{figure}
The shock still exists as long as Eq. (1) is verified. Combining with
 Eq. (7), Eq. (1) becomes:
\begin{eqnarray}
  \left.\frac{\partial^2 \widetilde{u}}{\partial
      \widetilde{x}^2}\right|_{x=0}=P(\widetilde{u})
      &=&49\widetilde{u}^3-84\widetilde{u}^2+(39+42\Pi)\widetilde{u}\nonumber\\
      & & -24\Pi+18\frac{\Pi}{\widetilde{R}_{\gamma\gamma}}-4=0
      \label{pu}
\end{eqnarray}
Since $P(\widetilde{u})$ is a third degree polynomial, it possesses in
general 3 real solutions which depends obviously on $\Pi$ and
$\widetilde{R}_{\gamma\gamma}$. They have been plotted in Fig. 2 in
function of $\Pi$ for $\widetilde{R}_{\gamma\gamma}$=50. Of course,
for $\Pi$ going to zero, the three branches of solution converge
respectively to the well-known results $\widetilde{u}$=1/7, 4/7 and 1
corresponding to the values without pair creation. By continuity, the
flow velocity at the shock location will follow the second branch
noted $\widetilde{u}_{shock}$ on the figure (plotted in
\begin{figure}[h!]
\plotfiddle{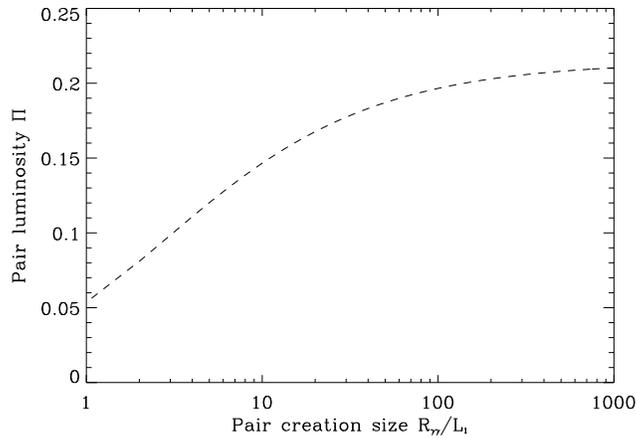}{5cm}{0}{50}{50}{-150}{-185}
\caption{Curve $\Pi_{max}$ vs. $\widetilde{R}_{\gamma\gamma}$. We see
that $\Pi_{max}$ is upperlimited by $\sim 0.2$.
\label{a_lrelation}}
\end{figure}
dashed line). {\bf It appears that for a given value of
$\widetilde{R}_{\gamma\gamma}$ it exists a maximal value $\Pi_{max}$
of $\Pi$ above which there is only one real solution which still
verifies Eq. (11). The unphysical discontinuity of the flow velocity
at the shock location for $\Pi=\Pi_{max}$ means simply that the shock
can not exist anymore.}\\

The transition between 3 to 1 real solution of Eq. (11) happens when
the following conditions are satisfied:
\begin{eqnarray}
  P(\widetilde{u}) &=& 0\\
  P'(\widetilde{u}) &=& 0
\end{eqnarray}
\begin{figure}[h!]
\plotfiddle{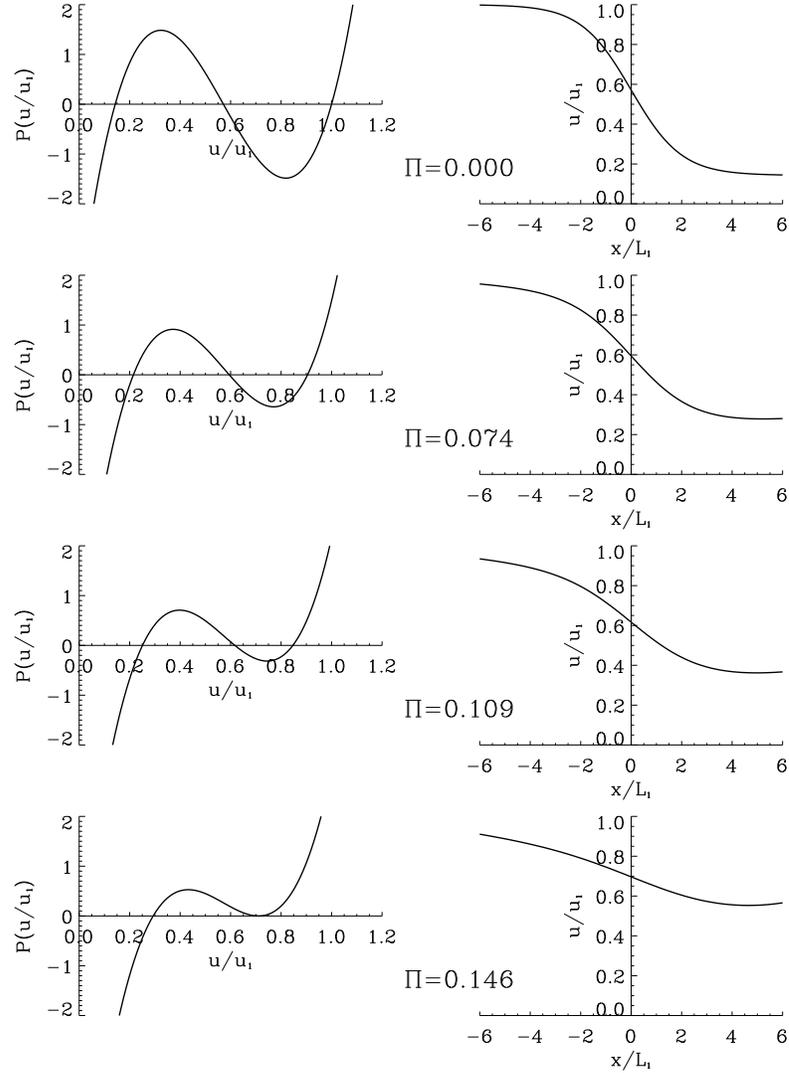}{15cm}{0}{60}{60}{-150}{-50}
\caption{disappearance of a shock due to pair creation. We have taken
$\widetilde{R}_{\gamma\gamma}$=10 meaning that $\Pi$ must be smaller
than $\sim 0.15$ (cf. Fig. 3).  \label{profchoc}}
\end{figure}
The resolution of this system of equations gives thus a relation
between $\Pi_{max}$ and $\widetilde{R}_{\gamma\gamma}$:
\begin{equation}
  \Pi_{max}=\frac{3}{14}-
  \frac{1}{14}\left(63\frac{\Pi_{max}}{\widetilde{R}_{\gamma\gamma}}
	\right)^{2/3} 
  \label{condnec}
\end{equation}
We have reported the corresponding function
$\Pi_{max}(\widetilde{R}_{\gamma\gamma})$ in Fig. 3. We see that
$\Pi_{max}$ is a increasing function of $\widetilde{R}_{\gamma\gamma}$
meaning that the larger the pair creation region the larger the pair
power we need to kill the shock. However, we see from Eq. (14) that
$\Pi_{max}$ is necessarily smaller than 3/14$\simeq$ 0.20 meaning that
{\bf at most 20\% of the kinetic energy flux of the upstream flow
transformed in pairs is sufficient to suppress the shock
discontinuity}.\\

We have reported in Fig. 4, different shape of the polynomial
$P(\widetilde{u})$ and the corresponding flow velocity profiles
obtained numerically by solving Eq. (7) for different values of $\Pi$
(fixing $\widetilde{R}_{\gamma\gamma}$ to 10). We have also plotted in
Fig. 5, the variation of the compression ratio $r$ (as defined in
section 2.1) in function of the pair creation rate $\Pi$. As expected,
$r$ converge to $\sim$~1 when $\Pi$ increases meaning that the
acceleration becomes less and less efficient.

\begin{figure}[h!]
\plotfiddle{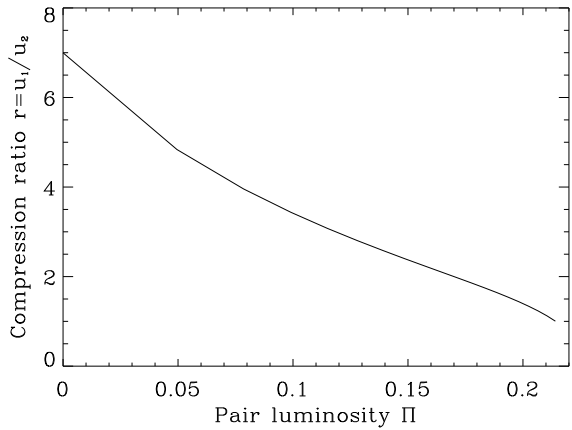}{5cm}{0}{50}{50}{-150}{-185}
\caption{Variation of the compression ratio $r$ in function of the
pair luminosity $\Pi$ for $\widetilde{R}_{\gamma\gamma}=1000$.}
\end{figure}

\section{Stationary states}
Up to now, we have supposed that the deformation of the flow profile,
due to the pair pressure, does not modify the pair creation rate
itself.  This is a crude assumption since a change in the velocity
profile of the flow will modify the distribution function of the
accelerated particles in such a way that the number of particles
enable to trigger the pair creation process (i.e.  particles with
$\gamma\ge\gamma_{th}$) may change, modifying consequently the pair
creation rate.\\ From the previous section we have seen that the
compression ratio $r$ decreases when the pair creation rate raises
(cf. Fig. 5) but reversely if $r$ decreases, the spectral index will
increase and the number of high energy particles will decrease, the
final effect being a decrease of the pair creation rate. Consequently,
we may expect the system to reach, in some conditions, stationary
states where hydrodynamics and pair creation effects balance.\\

\subsection{The parameter space}
We have studied the stationary states of our toy model by solving
self-consistently for the particle distribution function $n(\gamma)$
and the flow velocity profile $u(x)$, the two function $n(\gamma)$ and
$u(x)$ being linked through the pair pressure. In stationnary states,
the system depends on six different parameters: $\gamma_{min}$, the
minimal Lorentz factor for the particles to be accelerated in the
shock (cf. section 2.2), $\epsilon_s$, the external soft photon energy
(in unit of $m_ec^2$), $u_1$, the upstream flow velocity,
$\widetilde{R}$ the transverse size of the shock (in unit of the
diffusion length $L$), $l_s$, the compactness of the external soft
photon field (cf. section 2.3) and $l_{kin}$ the kinetic compactness
defined as $\displaystyle l_{kin}=\frac{(\rho u_1^3\pi
R^2)\sigma_T}{4\pi Rm_ec^3}$. The larger $l_s$ and the larger the
cooling of the particles, whereas the larger $l_{kin}$ and the larger
the kinetic energy of the upstream flow.\\

The system is solvable in only some part of the parameter space. It
always possesses two solutions: a ``pair dominated'' (large pair
luminosity $\Pi$, small compression ratio $r$) and a ``pair free'' one
(small $\Pi$, large $r$). The latter connects to the trivial solution of
the problem i.e. $\Pi =0$.\\

\subsection{High energy spectra}
For a given set of parameters, the system may reach stationary states
characterized by a compression ratio $r$. The spectral index and cut--off
of the particle distribution function (Eq. (4)) are then given by Eq. (2)
and (3) respectively. Since we have assumed that these particles are
cooled by Inverse Compton effect and that their distribution function
follows Eq. (4), the emitted energy spectrum is characterized by a
cut-off power law shape $\displaystyle F_{E}\propto E^{-\alpha}\exp
\left[-\left(\frac{E}{E_c}\right)^{\frac{1}{2}}\right]$ where $\alpha$
and $E_c$ are simple functions of $s$ and $\gamma_c$ (Rybicki \&
Lightman, 1979):
\begin{eqnarray}
\alpha&=&\frac{s}{2}\\
E_c&\simeq&\frac{4}{3}\gamma_c^2\epsilon_sm_ec^2
\end{eqnarray}
\begin{figure}[h!]
\plottwo{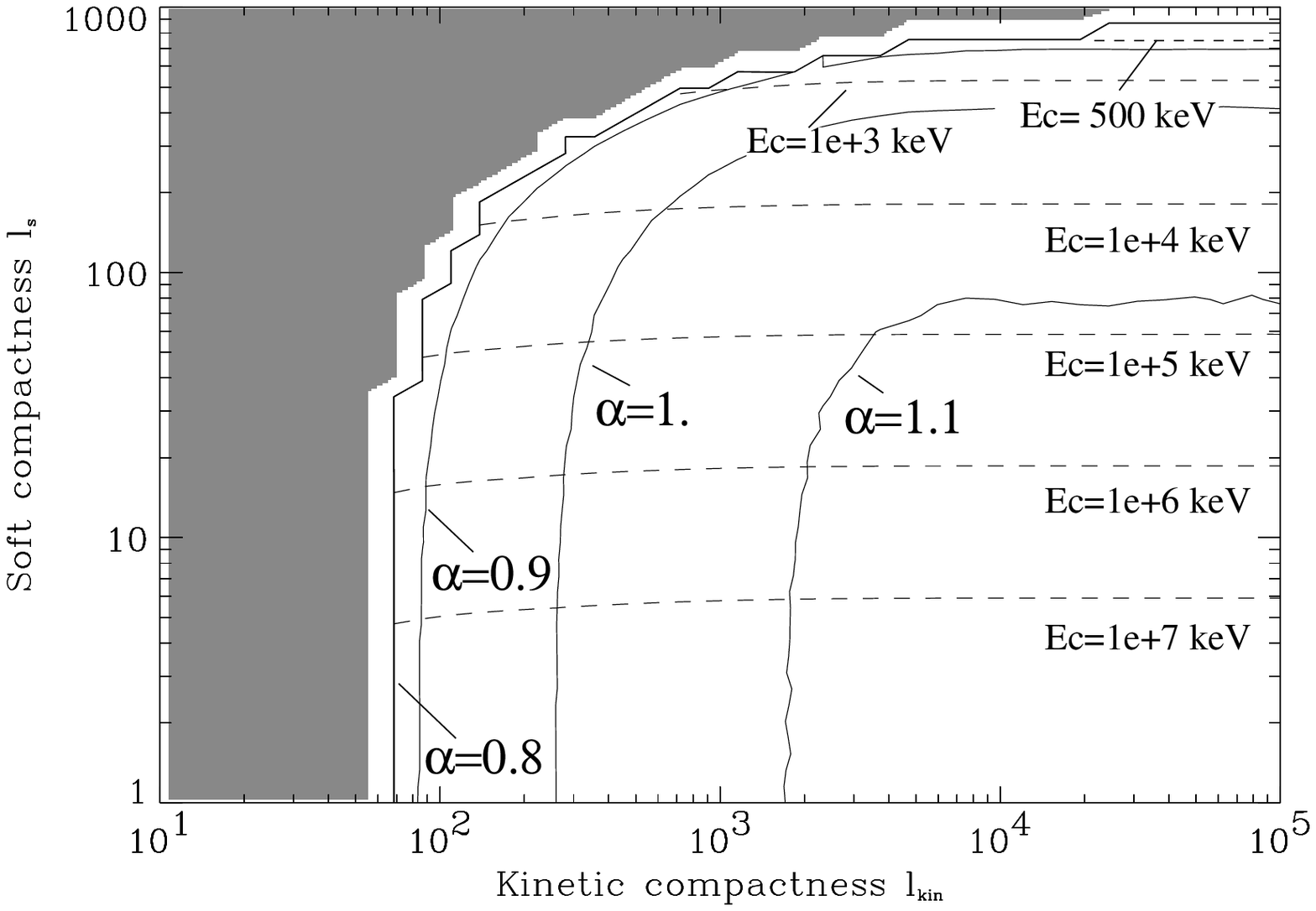}{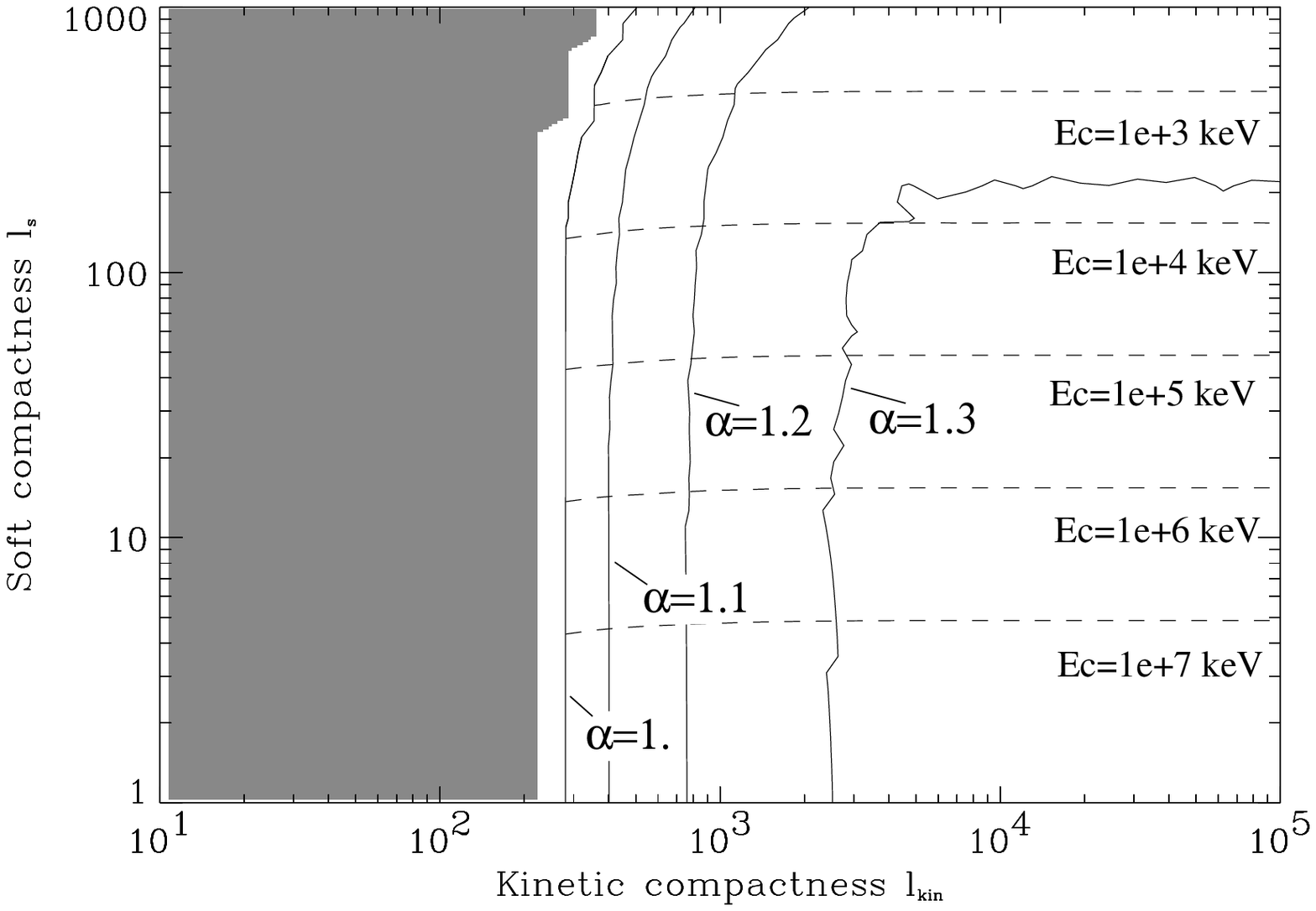}
\caption{Contour plots of the spectral index $\alpha$ (solid lines)
and the high energy cut-off $E_c$ in keV (dashed lines) of the
emitted spectrum in the ($l_s$, $l_{kin}$) space. The soft photon
energy is equal to 10 eV and 600 eV in the left and right plot
respectively. The other parameters have been fixed to
$\widetilde{R}=10^8$, $\gamma_{min}=10$ and $u_1/c=0.1$.In the grayed
region there is no stationnary state solution.}
\end{figure}

We have reported in Fig. 6 the contour plot of $\alpha(l_s,l_{kin})$ and
$E_c(l_s,l_{kin})$ for two values of $\epsilon_s$ (10 and 600 eV which
are representative of the typical values of soft photons emitted by an
accretion disk around a supermassive and stellar mass black hole
respectively). The other parameters have been fixed to
$\widetilde{R}=10^8$, $\gamma_{min}=10$ and $u_1/c= 0.1$. The contours of
$\alpha$ and $E_c$ keep roughly the same shape but cover a different
region of the parameter space for different parameter sets. In each
figure, we can see that:
\begin{itemize}
\item The spectral index does not strongly vary between the harder and
the softer spectra ($\Delta\alpha\simeq0.3$). It reaches an asymptotic
plateau for low $l_s$ and high $l_{kin}$. In these conditions, both pair
density and high energy cut--off are large so that the pair creation
process is saturated.  Concerning $E_c$, following Eqs. (3) and (16), it
is inversely proportional to $l_s^2$.
\item The harder spectra are obtained for large values of $l_s$ or small
values of $l_{kin}$. In both case, the pair efficiency decreases either
due to a small value of $E_c$ or a low particle density, that is a low
pair production optical depth. To keep the equilibrium between the pair
creation effects and the hydrodynamic of the flow, the system has to
reach harder spectra to compensate this decrease of the pair creation
rate.
\item For small $l_s$, $\gamma_c$ is very large ($\gg \gamma_{th}$) and
its value becomes immaterial.  Consequently, $\Pi$ and $r$, and thus
$\alpha$ become independent of $l_s$
\item For large $l_{kin}$, the pair density must be also high to
efficiently modify the hydrodynamical profile. The pair creation process
is thus saturated meaning that the pair creation rate grows linearly with
$l_{kin}$, i.e. $\Pi$, $r$ and thus $\alpha$ become independent of
$l_{kin}$.
\item For to large $l_s$ or to small $l_{kin}$, the system cannot reach
sufficient hard states to keep in equilibrium and no high pair density
stationary state can exist any more. The system can only be in the
trivial pair free state (i.e. $r=7$ and $\Pi=0$).
\end{itemize}
We interpret the differences between the two plots of Fig. 6 as
follows. The increase of $\epsilon_s$ favors the pair creation
process. Thus, for given values of $l_s$ and $l_{kin}$, the spectral
index is larger. The high energy cut--off $E_c$ increases mainly because
of its dependence on $\epsilon_s$ (cf. Eq. (16)), $\gamma_c$ keeping
roughly constant (cf. Eq. (3)).

\subsection{Annihilation line}
The presence of pairs should give a signature as an annihilation feature
at $\sim$ 511 keV. We will show that this feature is not expected to be
strong in our model. Here, we have supposed the existence in the shock
region of pre--accelerating processes bringing leptons to the sufficient
energy (i.e. $\gamma > \gamma_{min}$ with $\gamma_{min}$ of the order of
a few, cf section 2.2) for resonant scattering off magnetic
disturbances. Since the annihilation process occurs mainly at low energy,
i.e. for particles with Lorentz factor $\gamma\simeq 1$ (cf. Coppi \&
Blandford 1990), it occurs mainly far downstream, where the pairs created
in the shock can cool down. The annihilation line luminosity is thus at
most equal to the pair rest mass luminosity.  As shown in section 4, the
pair luminosity $\Pi$ is itself limited and is necessarily smaller than
$\sim$20\% of the X-ray/$\gamma$-ray luminosity , i.e. $\Pi<
\Pi_{max}\simeq 0.2$ (assuming that the total kinetic energy of the
upstream flow is transformed in radiation). Besides, for $\Pi\simeq 0.2$,
the compression factor is very small, of the order of unity, resulting in
a very steep X-ray spectra.  When hydrodynamics feedback is taken into
account, the pair luminosity may be well below this theoretical limit of
20\%. An X-ray spectral photon index $\sim$ 2 (as those generally seen in
Seyfert galaxies) requires a compression ratio $r\sim$ 3--4. Such values
of $r$ require values of $\Pi$ smaller than $\simeq$ 10\% (cf. Fig 5).\\

Assuming a steady state, pairs annihilate at the same rate as they are
produced. $\Pi/\gamma_{min}$ gives then an upper limit of the
annihilation line luminosity. We thus expect the luminosity of the
annihilation radiation to be smaller than few percent of the total high
energy radiation, which is quite compatible with the non observation of
strong annihilation lines in this class of Seyfert galaxies as shown by
the best upper limit observed in Seyfert galaxies with the
OSSE satellite (Johnson et al. 1997).\\

\subsection{Variability}
For some values of external parameters, as suggested by Fig. 6, no high
pair density solution can exist in stationary states. So only the pair
free solution (i,e, with $\Pi=0$) exists. The system is not expected to
be variable with a constant set of parameters and variability can only
occur with a variation of one of them. An interesting possibility would
be to consider a possible feedback of the relativistic plasma to the soft
compactness: in the reillumination models for instance (Collin, 1991;
Henri \& Petrucci, 1997), the soft photons are produced by the
reprocessing of the primary X-ray emission. An increase of the pair
plasma density will increase the X-ray illumination and thus the soft
compactness. Fig. 6 shows that in some cases, the change of $l_s$ make
the system switch to pair free solution which will stop the
reillumination and bring the system back to pair rich solutions.
Limiting cycles could thus occur. We intend to further investigate this
possible effect under astrophysically relevant conditions.

\section{Conclusion}
In the present paper, we have studied the effect of pair creation, via
high energy photon-photon interaction, on a shock structure, where the
high energy photons are produced via IC by the particles accelerated by
the shock itself. The problem is highly nonlinear since pairs can modify
the shock profile through their pressure and, mutually, a change of
the shock hydrodynamics can decrease or increase the pair production
rate.\\

We have shown that for a given size of the pair creation region, it exist
a maximal value of the pair creation rate above which the shock cannot
exist anymore.  When the hydrodynamical feedbacks on the pair creation
process are neglected, a pair power of at most 20\% of the upstream
kinetic power is sufficient to kill the shock. This constraint can fall
to few percents in stationary states where pair creation and
hydrodynamical effects balance.  We thus do not expect the presence of
strong annihilation lines. We also obtain spectral parameters in rough
agreement with the observations.\\

We suggest also a possible variability mechanism if the soft photon
compactness depends itself on the pair density of the hot plasma, such as
expected in reillumination models.\\

In the model presented here, the cooling of particles is due to the IC
process on external soft photons. However, particles may also cool on
soft photon they produce by synchrotron process when spiraling around the
magnetic field lines (the so-called synchrotron-self-compton process,
SSC). In this case the cooling will also depend on the particle
distribution function. We may expect that the addition of the SSC process
would allow to obtain stationary states with harder spectra than those we
obtained here, since the additional synchrotron cooling would be
compensate by a stronger acceleration, i.e. a larger compression ratio
$r$. The detailed study of this problem is left to future work.

{\it Acknowledgements:} POP acknowledges a grant of the European
Commission under contract number ERBFMRX-CT98-0195 (TMR network
"Accretion onto black holes, compact stars and protostars").


\end{document}